\documentclass[aps,preprint,superscriptaddress]{revtex4}
\usepackage{graphicx}
\usepackage{amssymb}
\usepackage{epsfig}

\begin{document}

\title{Dynamical estimates of chaotic systems from Poincar\'e recurrences}

\author{M.  S.  Baptista}
\affiliation{Institute for Complex
Systems and Mathematical Biology, King's College, University of
Aberdeen, AB24 3UE Aberdeen, United Kingdom}
\affiliation{Centro de Matem\'atica da Universidade do Porto, Rua do Campo
Alegre 687, 4169-007 Porto, Portugal}
\author{Dariel M. Maranh\~ao}
\affiliation{Institute of Physics, University of S\~ao Paulo,
Rua do Mat\~ao, Travessa R, 187, 05508-090 SP, Brasil}, 
\author{J. C. Sartorelli}
\affiliation{Institute of Physics, University of S\~ao Paulo,
Rua do Mat\~ao, Travessa R, 187, 05508-090 SP, Brasil}


\begin{abstract}
  We show a function that fits well the probability density of return
  times between two consecutive visits of a chaotic trajectory to
  finite size regions in phase space. It deviates from the exponential
  statistics by a small power-law term, a term that represents the
  deterministic manifestation of the dynamics. We also show how one
  can quickly and easily estimate the Kolmogorov-Sinai entropy and the
  short-term correlation function by realizing observations of high
  probable returns. Our analyzes are performed numerically in the
  H\'enon map and experimentally in a Chua's circuit. Finally, we
  discuss how our approach can be used to treat data coming from
  experimental complex systems and for technological applications.
\end{abstract}

\maketitle

{\bf Observing how long a dynamical system takes to return to some
  state is one of the simplest ways to model and quantify its dynamics
  from data series. In this work, we describe a simple way to extract
  some relevant invariant quantities of a chaotic system by using
  recurrence times, in particular Poincar\'e recurrences that measure
  the time interval for a system to return to a configuration close to
  its initial state.  Part of this work is dedicated to apply the
  theoretical results proposed by [Pinto {\it et al.}
  arxiv:0908.4575] to calculate the Kolmogorov-Sinai entropy and the
  decay of correlation of the experimental Chua's circuit when the
  returns are measured in ``large'' size regions in phase space.
  Another part is dedicated to study how the deviation (from the
  exponential form) of the density of the first Poincar\'e returns can
  be used to detect deterministic manifestations in chaotic systems.
  Finally, we discuss how our approach can be used to treat data
  coming from experimental complex systems and for technological
  applications.}

\section{Introduction}

Chaotic systems have simultaneously a stochastic \cite{stocastic} and
a deterministic dynamical characters \cite{dinamica}. Single
trajectories are predictable (deterministic) for a short-term
evolution and unpredictable for a long-term evolution
(stochastic). While the stochastic character is associated with an
exponential decay of correlations and information about the actual
state is rapidly lost as the trajectory evolves, the deterministic
character is associated to a power-law decay of correlations and
information about the actual state is lost as the system is evolved,
but in a rather slower fashion.

A relevant question that arises when dealing with data coming from
complex systems is whether manifestations of chaotic behavior can be
detected in the data \cite{mamen} so that one can construct
deterministic models. One wishes to come up with a dynamical
description of the data, but due to the sensitivity dependence on
initial conditions of chaotic systems one is prompt to adopt a
probabilistic approach in order to reveal the underlying dynamics of
the system from statistical averages
\cite{nicolis}. A promising tool of analysis is provided by the
statistics of the Poincar\'e recurrence time (PRT)
\cite{poincare} which measures the time interval between two 
consecutive visits of a trajectory to a finite size region in phase space.

Many relevant quantifiers of low-dimensional chaotic systems can be
obtained by the statistical properties of the PRTs. 
The purpose of the present work is to apply some results from
Ref. \cite{paulo} and to propose other theoretical approaches to
easily identify deterministic and stochastic manifestations in
dissipative strongly chaotic systems by using the PRTs of chaotic
trajectories to regions of finite size
\cite{baptista}, considering only short return times. 
These later two conditions constraining our analyzes are devoted to
suitably apply our approaches to realistic physical situations: the
resolution to measure returns as well as the time frame to realize the
experiment is finite.

By chaotic systems, we mean non-uniformly hyperbolic systems.  We
focus the analyzes on the H\'enon map and on an experiment, the Chua's
circuit, but we also present results coming from the logistic map.

We first present the theoretical framework to be used in Secs.
\ref{secao_probabilidade},\ref{secao_parametros},\ref{secao_lyapunov}, and
\ref{secao_correlacao}. 
Then, in Sec. \ref{secao_aplicacao}, we apply this approach to analyze
the H\'enon map and the experimental
Chua's circuit. We also work with the
Logistic map in Appendix \ref{apendice}.

In Sec. \ref{secao_probabilidade}, we present a continuous function,
$\rho_F$ that fits well the probability distribution for the PRTs to
regions of finite size. The many parameters of $\rho_F$ are
theoretically estimated in Sec.
\ref{secao_parametros}. Further, we show how to use  these parameters 
to quantify how stochastic or deterministic the considered system is,
under the perspective of the PRTs.  Concerning the coefficient of the
power-law term (responsible for the deviation of $\rho_F$ from the
Poisson distribution), a first approximation furnishes that it is
inversely proportional to the average return time. The longer
(shorter) is the average return time, the smaller (larger) the
power-law term, and the more stochastic (more deterministic) the PRTs.

In Sec. \ref{secao_lyapunov}, we succinctly describe how to calculate
the Kolmogorov-Sinai (KS) entropy, denoted by $H_{KS}$, in terms of
the number of unstable periodic orbits (UPOs) and of the frequency with which the PRT
happen. Then, we explain how $H_{KS}$ can be calculated in typical
physical situations, when the information about the UPOs is unknown
and the only available information is the frequency with which a high
probable PRT with short length happens.  Notice that this frequency
can be easily measured even if only a few return times are observed
because faster returns are more probable.

Alternative methods to calculate the correlation entropy, $K_2$, a
lower bound of $H_{KS}$, and to calculate $H_{KS}$ from time series
were proposed in Refs. \cite{grassberger,cohen}. In
Ref. \cite{grassberger} $K_2$ is calculated from the correlation decay
and in Ref. \cite{cohen} by the determination of a generating
partition of phase space that preserves the value of the entropy. But
while the method in Ref. \cite{grassberger} unavoidably suffers from
the same difficulties found in the proper calculation of the fractal
dimensions from data sets, the method in Ref. \cite{cohen} requires
the knowledge of the generating partitions, information that is not
trivial to be extracted from complex data. The advantage of the
theoretical approach used is its simplicity. As one can see in
Eq. (\ref{upper_bound3}), the only information required is the
frequency with which high probable PRTs happen in some regions of the
phase space.

In Sec. \ref{secao_correlacao}, we show how the PRTs can be used to
calculate the correlation function for short-term returns and an upper
bound for long-term returns, a function that indicates how much the
future returns are related to the past returns to a finite size
region.

Other methods to calculate the correlation function from the PRTs were
proposed in Refs. \cite{correlation2,young}. One of the advantages of
the proposed theoretical approach is that we could show, in a very
trivial way, that if the distribution of the PRTs has an exponential
decay then the correlation function also decays exponentially, a
result rigorously demonstrated in \cite{young}.

Finally, in Sec. \ref{secao_conclusao}, we present our conclusions and
discuss how this approach can be used in an integrated way to
characterize experimental data coming from complex systems.

\section{The probability distribution of the PRTs}\label{secao_probabilidade}
 
In recent works \cite{eduardo,baptista:2005}, it was shown that the
fitted probability density function $\rho_F(\tau,{\mathcal{B}})$, of a
series of Poincar\'e return times $\tau_n$ ($n=1,\ldots,L$) to a box
${\mathcal{B}}$ of equal {\it finite} sides $2\epsilon$ \cite{return},
of typical trajectories in a non-uniform hyperbolic attractor on
$\mathcal{R}^2$, deviates from the exponential law
$\rho_P(\tau,{\mathcal{B}})=\mu({\mathcal{B}})e^{-\tau
\mu({\mathcal{B}})}$ 
if the box ${\mathcal{B}}$ is placed on some special region of the
phase space, where $\mu({\mathcal{B}})$ is the probability measure
inside it. This quantity measures the frequency with which a chaotic
trajectory visits ${\mathcal{B}}$.

The function $\rho_P(\tau,{\mathcal{B}})$ describes well the discrete probability distribution function (PDF),
denoted by $\rho(\tau, {\mathcal{B}})$, in Axiom-A systems and some
classes of non-uniformly hyperbolic systems with exponential decay of
correlations (=chaos with strong mixing properties)
\cite{hirata}, for arbitrarily small intervals ($\epsilon\rightarrow 0$).
In Ref. \cite{baptista:2005}, it was hypothesized that for 
$\tau>\tau_{UPO}^{min}$, we have 
$\rho(\tau,{\mathcal{B}}) \rightarrow  {\rho}_F^{\prime}(\tau,{\mathcal{B}}) 
$ with
\begin{equation}
{\rho}_F^{\prime}(\tau,{\mathcal{B}})  = \beta e{^{(-\alpha\tau)}}, 
\label{return_dist}
\end{equation}
\noindent
where $\tau>\tau_{UPO}^{min}$ represents the PRT for which the
distribution $\rho(\tau,{\mathcal{B}})$ becomes approximately
continuous and $\beta=\beta_{R}+\alpha$.  It was assumed that $\beta_R
e{^{(-\alpha\tau)}}$ describes an effective PDF associated mostly with
the return of trajectories along the unstable manifold of UPOs outside
the box ${\mathcal{B}}$.  It is the result of the non-hyperbolic
nature of the dynamics inside the box, whose probability measure
suffers the influence of non-local UPOs. The term $\alpha
e{^{(-\alpha\tau)}}$ describes the hyperbolic nature of the dynamics
inside the box, the return of trajectories to the box associated with
the local dynamics provided by the UPOs inside the interval.

The coefficients of $\rho_F^{\prime}(\tau, {\mathcal{B}})$ are obtained by the
least square fitting method which minimizes the error between
$\rho_F^{\prime}(\tau, {\mathcal{B}})$ and $\rho(\tau, {\mathcal{B}})$: {\small
\begin{equation}
E(\rho_F^{\prime}-\rho)=(\rho_F^{\prime}-\rho)^2.
\label{error}
\end{equation}
} 
 
In addition, assuming that Eq. (\ref{return_dist}) describes well
$\rho(\tau, {\mathcal{B}})$, then the following equations should be
simultaneously satisfied {\small
\begin{eqnarray}
\sum_{\tau_{min}}^{\tau^{min}_{UPO}} \rho(\tau,{\mathcal{B}})
+ \int_{\tau^{min}_{UPO}}^{\tau_{max}} \rho_F^{\prime}(\tau,{\mathcal{B}}) d\tau&=&1
\label{cond1} \\
\sum_{\tau_{min}}^{\tau^{min}_{UPO}} \rho(\tau,{\mathcal{B}}) \times \tau +
\int_{\tau^{min}_{UPO}}^{\tau_{max}} \rho_F^{\prime}(\tau,{\mathcal{B}}) \tau d\tau&=&\langle \tau \rangle
\label{cond2}
\end{eqnarray}
} 
\noindent
by considering that $\rho(\tau,{\mathcal{B}})$ can be broken in two
terms, one that describes its discrete nature, the probability of
finding a PRT of length $\tau<\tau^{min}_{UPO}$, and another that
describes its continuous nature. 

Denoting $\tau_{min}$ as to be the PRT with the minimum length
[$\min{(\tau_n)}$] and $\langle \tau
\rangle=1/L\sum_{n=1}^L \tau_n$, and assuming that 
$\tau_{min}=\tau^{min}_{UPO}$ then $\alpha \cong (\langle \tau
\rangle - \tau_{min})^{-1}$ and $\beta_{R} \cong \alpha [e^{(\alpha \tau_{min})} -1]$, 
for sufficiently small $\epsilon$ so that the terms that contain
$\max{(\tau_n)}$, regarded as $\tau_{max}$, can be neglected.  So, for
finite $\langle \tau
\rangle$, $\beta > \alpha$. However, Eq. (\ref{return_dist}) does not
seem to completely capture the nature of $\rho(\tau,{\mathcal{B}})$
for finite size intervals.  Satisfying conditions (\ref{cond1}) and
(\ref{cond2}) do not necessarily minimizes the error in
(\ref{error}), and the contrary also does not apply.  Such
disagreement becomes stronger in boxes centered close to homoclinic
tangencies.  By fitting $\rho(\tau,{\mathcal{B}})$ by a
function of the type in Eq.  (\ref{return_dist}) might produce $\beta
\leq \alpha$, which disagrees with the inequality $\beta > \alpha$ 
that validates Eq. (\ref{return_dist}).  For the here considered
dynamical systems for parameters far away from intermittent behavior,
and data coming from plasma turbulence and stock market
\cite{baptista}, one finds often that $\alpha
\cong \beta$, a consequence of $\rho(\tau,{\mathcal{B}})$ be greater than 
$\rho_F^{\prime}(\tau,{\mathcal{B}})$, for $\tau<
\langle \tau \rangle$, and that $\rho(\tau,{\mathcal{B}})<
\rho_F^{\prime}(\tau,{\mathcal{B}})$, for $\tau >
\langle \tau \rangle$.  
These facts suggest that for systems with a dynamics similar to the
H\'enon map, the small term $\beta_{R}$ considered to be constant in
Ref. \cite{baptista:2005} is in fact a function that we hypothesize to
be a power-law with respect to $\tau^{\prime}$. That lead us to
\begin{equation}
\rho_F(\tau^{\prime},{\mathcal{B}}) = 
\left(\beta^*+c\left[1-\left(\frac{\tau^{\prime}}{\langle \tau_e \rangle}\right)^{\gamma}\right]\right)e^{(-\alpha\tau^{\prime})}
\label{return_dist1}
\end{equation}
\noindent 
where 
\begin{eqnarray}
\tau^{\prime} &=& \tau-\tau_{UPO}^{min}+1 \label{tau_prime} \\
\langle \tau_e \rangle &=&
\int_{\tau^{min}_{UPO}}^{\tau_{max}}
\rho(\tau^{\prime},{\mathcal{B}}) \tau^{\prime} d\tau^{\prime} \label{tau_e}
\end{eqnarray}
\noindent
and $\gamma$ is a positive small value. Notice that the use of
$\tau^{\prime}$ and $\langle \tau_e \rangle$ is only an artifact to
simplify our further approximations and also to simplify any possible
nonlinear fitting that we might make concerning the real PDF.

The power-law term $\beta_R=c[1-(\frac{\tau^{\prime}}{\langle \tau_e
\rangle})^{\gamma}]$ shows that the considered systems for which 
this distribution holds have a return time distribution whose decay is not
characterized by either a power-law or exponential decay but by both.  It
appears as a combined effect of the finiteness of the box, the expanding
factor nearby low-periodic UPOs along the unstable direction, and the
existence of homoclinic tangencies, which gives the non-uniformly hyperbolic
character of the H\'enon map and the Chua's circuit.

A large value of $\gamma$ (which implies in a large $\beta_R$)
indicates that there is a large contribution to the measure inside the
box due to UPOs that are outside the box. Two things contribute for a
large value of $\gamma$: the size and location of the box. The larger
the size and the level of non-hyperbolicity of the box are, the larger
the value of $\gamma$ is. We find larger values of $\gamma$ when we
measure PRTs in boxes placed close to homoclinic tangencies. At such a
case, trajectory points that have returned once to the box after $P$
iterations keep consecutively returning to the box after $P$
iterations. The trajectory is no longer under the influence of the
linear expanding character of a period-$P$ UPO (along the direction of
the unstable eigenvector) but under the influence of the non-linear
character of the unstable manifold of the UPO located outside the box.

\begin{figure}[!htb]
\centerline{\hbox{\psfig{file=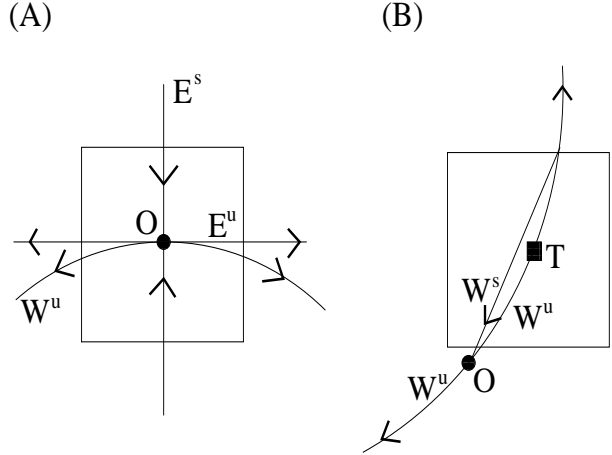,height=6cm,width=8cm}}}
\caption{A sketch of the structure of the manifolds and eigenvectors 
of UPOs (represented by full circles and the letter {\bf O}) inside (A) or
outside (B) a finite size box, where the PRTs are being measured. In (A), the
box is far away from homoclinic tangencies, and in (B) the box is in the
neighborhood of a homoclinic tangency (represented by the full square and the
letter {\bf T}).}
\label{fig_manifolds}
\end{figure}

A sketch of these ideas is depicted in Fig.
\ref{fig_manifolds}. In (A), we represent a box that is centered on an
UPO (in the figure represented by a full circle and the letter {\bf
O}). The box is sufficiently large so that the escape of trajectories
from the box which are associated with this UPO happens no longer
along the unstable eigenvector ($E^u$) but along the unstable manifold
($W^u$). The smaller is the period of an UPO, the smaller is the
eigenvalue of the unstable eigenvector. So, the escape of the
trajectory along the unstable manifold is usually expected to happen
for a low-period UPO. In (B), we represent a box centered in a
homoclinic tangent (in the figure represented by a full square and the
letter {\bf T}). In the neighborhood of homoclinic tangencies exist
low-period UPOs whose unstable and stable manifold happens to be
almost parallel, forming a trapping region.  In these trapping
regions, the trajectory returns to the box along $W^{u}$ and
eventually escapes the box, but it may return to it along the stable
manifold $W^s$ of the UPO outside the box. For short returns, these
confined trajectories contribute positively to
$\rho(\tau,{\mathcal{B}})$, increasing $\mu({\mathcal{B}})$.

We can classify regions ${\mathcal{B}}$ by the way returns happen inside it:
{\it type I} are the regions whose returns are associated with higher-period
UPOs, here conveniently regarded as the hyperbolic ones. {\it type II and III}
are the ones associated with lower-period UPOs, here conveniently regarded as
the non-hyperbolic regions.  In these regions, the deviation of $\rho_F(\tau,{\mathcal{B}})$ to
the exponential is large, meaning a large $\gamma$. For the {\it type II}
regions, the unusual returns that happen inside the interval are associated to
UPOs outside it. Such behavior is a consequence of the existence of homoclinic
tangencies
\cite{baptista:2005} inside the region.   For the {\it type III} regions, 
the unusual returns are associated with lower-period UPOs inside them.
\cite{baptista:2005}. 
When there is an UPO with low period $P$ inside the box,
$\rho(\tau=P)$ is no longer exactly equal to $\mu_{NR}$ [given by
Eq. (\ref{mu_NR1})] due to the fact that linearization around this
low-period UPO does not provide the measure due to this UPO inside the
box.
 
\section{Estimation of $\alpha$, $\beta^*$, $\tau_{UPO}^{min}$, $\omega$, and $\gamma$}\label{secao_parametros}

To estimate $\alpha$ and $\beta^*$, we make $\gamma$=0 (treating the
power-law term as a perturbation), and place
$\rho_F(\tau,{\mathcal{B}})$ into Eqs. (\ref{cond1}) and
(\ref{cond2}). We arrive that {\small
\begin{eqnarray}
\beta^*&=& \frac{\rho_e^2}{\langle \tau_e \rangle} \label{estima_beta} \\
\alpha &=& \frac{\rho_e}{\langle \tau_e \rangle} \label{estima_alpha}
\end{eqnarray}}
\noindent
where $\rho_e({\mathcal{B}})=\int_{\tau_{UPO}^{min}}^{\tau_{max}}
\rho(\tau,{\mathcal{B}})d\tau$ and $\langle \tau_e \rangle$ defined as
in Eq. (\ref{tau_e}).

As discussed in \cite{baptista:2005}, the exponential part of the distribution
reflects the stochastic nature of the returns associated with UPOs inside
${\mathcal{B}}$. It is interesting to know how much the exponential term
$\beta^* \exp{[-\alpha \tau^{\prime}]}$ inside $\rho_F(\tau,{\mathcal{B}})$
deviates from the Poisson function
$\frac{\mu}{1-\mu}\exp{[\ln{(1-\mu)\tau^{\prime}}]}$, a function that
describes the probability with which returns happen if one considers
trajectories generated by uncorrelated processes. For that, we calculate
$\tilde{\mu}_1$ and $\tilde{\mu}_2$ such that {\small
\begin{equation}
\frac{\tilde{\mu}_1}{1-\tilde{\mu}_1} \exp{[\ln{(1-\tilde{\mu}_2)}
\tau^{\prime}]} \cong \beta^* \exp{[-\alpha \tau^{\prime}]}
\end{equation}}
\noindent
We arrive that $\tilde{\mu}_1 \cong \frac{\beta^*}{1+\beta^*}$ and $\tilde{\mu}_2
\cong
\alpha$. The continuous part of $\rho_F(\tau>\tau^{min}_{UPO})$ 
deviates from the Poisson distribution (for which
$\tilde{\mu}_1$=$\tilde{\mu}_2$) whenever $\tilde{\mu}_1 \neq
\tilde{\mu}_2$. Such
deviation becomes larger, the larger $|\mu - \alpha|$ ($=|\frac{\mu_e}{\rho_e} -
\frac{\rho_e}{\langle \tau_e \rangle}|$, where 
$\mu_e({\mathcal{B}})=\mu({\mathcal{B}})\rho_e({\mathcal{B}})$). The
larger $|\mu - \alpha|$ is, the larger $\tau_{UPO}^{min} - \tau_{min}$
(notice that $\mu = \alpha$ when $\tau_{min} = \tau_{UPO}^{min}$, what
happens when $\epsilon
\rightarrow 0$) and since $\alpha$ depends on $\tau_{UPO}^{min}$ 
it is reasonable to consider that the larger (the smaller)
\begin{equation}
\omega = (\tau_{UPO}^{min} - \tau_{min})
\label{dynamical_component}
\end{equation}
\noindent
is, the more correlated and deterministic (uncorrelated and stochastic) the
returns are. 

There are three approaches to estimate $\tau_{UPO}^{min}$. One is by just
inspecting when $\rho(\tau, {\mathcal{B}})$ presents a continuous exponential decay. The
other is by fitting $\rho(\tau, {\mathcal{B}})$ considering an exponential function of the
type ${\rho}^{\prime}_F(\tau, {\mathcal{B}})=\beta^* e^{-\alpha \tau}$, and $\tau_{UPO}^{min}$ is the
value for which $[{\rho}^{\prime}_F(\tau, {\mathcal{B}})-\rho(\tau, {\mathcal{B}})]$ becomes smaller than some $N/L$ ($N
\in
\mathbf{N}$, and $L$ is the number of PRT observed). The last approach is the
one that will be considered in this work due to its experimental
orientation. For that one has just to notice that as $\tau$ becomes large,
$\alpha$ as estimated by Eq. (\ref{estima_alpha}) approaches an asymptotic
value close to the value obtained by fitting $\rho(\tau,{\mathcal{B}})$ using the
exponential function ${\rho}^{\prime}_F(\tau,{\mathcal{B}})$.

To estimate $\gamma$, we
consider that $\tau_{max}-\tau_{UPO}^{min} >>1$, and use that  
$[1-(\frac{\tau^{\prime}}{\langle \tau_e \rangle})^{\gamma}] \cong
\gamma\ln{(\frac{\langle \tau_e \rangle}{\tau^{\prime}})}$, $\alpha \approx
\beta^*$, and $c \cong 1$ (so in the limit of $\epsilon \rightarrow 0$,
$\rho_F(\tau^{\prime},{\mathcal{B}}) \rightarrow \beta^* e^{-\alpha
\tau^{\prime}}$). Then, from Eq. (\ref{cond1}) we arrive that
\begin{equation}
\gamma \propto \frac{\tau_{min}}{\log{(\langle \tau \rangle)}\langle \tau \rangle^2}
\label{gama}
\end{equation}
\noindent
using that $\max{(\tau^{\prime}_{max})} \cong \tau_{max} >>1$. By
making $\alpha
\approx \beta^*$, we assume that the deviation of $\rho_F$ from the 
exponential law ($\alpha e^{-\alpha \tau}$) is exclusively provided by
the term $\beta_R$. In other words, the dynamical character of the
system is provided by $\beta_R$. Therefore, manifestations of the
deterministic behavior are more evident when
$\frac{\tau^{\prime}}{\langle
\tau_e \rangle}$ is maximal and that happens for when
$\tau^{\prime}=1$, what also means when $\tau=\tau_{min}$. 

To simplify the presentation of our results, we rewrite 
Eq. (\ref{gama}) as 
\begin{equation}
\gamma \propto \langle \tau \rangle^{\theta}
\label{teta}
\end{equation}

\section{Kolmogorov-Sinai entropy}\label{secao_lyapunov}

Now, we describe how the Kolmogorov-Sinai (KS) entropy can be written
in terms of the PDF. Then, we apply this formalism to a typical
physical situation when UPOs cannot be calculated and information
about them is unknown.

Firstly, remind that by the Pesin's equality
\cite{eckmann}, the KS-entropy equals the sum of the positive Lyapunov
exponents.

For uniformly hyperbolic systems \cite{paulo} (see Appendix), it is
valid to write that \begin{equation} \rho(\tau,{\mathcal{B}}) \cong
\frac{N_{NR}(\tau,{\mathcal{B}})}{N(\tau)} \label{rho_N}
\end{equation} \noindent where $N_{NR}(\tau,{\mathcal{B}})$ and
$N(\tau)$ represent the number of non-recurrent UPOs with period
$\tau$ inside ${\mathcal{B}}$ and $N(\tau)$ the total number of
different UPOs of period $\tau$ embedded in the chaotic attractor. A
non-recurrent UPO inside the region ${\mathcal{B}}$ is an UPO that
visits this region only once.

Also, for this class of systems, we can write that
\begin{equation}
N(\tau) = C \exp{^{\tau \times H_{KS}}}
\label{numeroUPO}
\end{equation}
\noindent
where $H_{KS}$ represents the Kolmogorov-Sinai entropy and $C$ is a
positive constant.

Then, placing Eq. (\ref{rho_N}) in (\ref{numeroUPO}), we have that
$H_{KS} =
-\frac{C}{\tau}+\frac{1}{\tau}\log{\left[\frac{N_{NR}(\tau,{\mathcal{B}})}{\rho(\tau,{\mathcal{B}})}\right]}$,
in the limit of $\tau \rightarrow \infty$. Then, for a finite $\tau$
we have that

\begin{equation}
H_{KS}(\tau,{\mathcal{B}}) \leq
\frac{1}{\tau}\log{\left[\frac{N_{NR}(\tau,{\mathcal{B}})}{\rho(\tau,{\mathcal{B}})}\right]}
\label{upper_bound}
\end{equation}

As shown in Appendix \ref{apendice}, it is reasonable to consider in
Eq. (\ref{upper_bound}) that $N_{NR}=1$. Even if there are more than
one non-recurrent UPO inside the region where the returns are being
measured, making $N_{NR}=1$ in Eq. (\ref{upper_bound}) will not
largely affect the estimation provided by this equation.  So, to
estimate the Lyapunov exponent of experimental systems, as the Chua's
circuit, we will consider that
\begin{equation}
H_{KS} (\tau,{\mathcal{B}}) \leq
\frac{1}{\tau}\log{\left[\frac{1}{\rho(\tau,{\mathcal{B}})}\right]}
\label{upper_bound1}
\end{equation}
\noindent
This equation offers a way to estimate the KS-entropy of a chaotic
system without having to calculate UPOs, a very difficult task.

Equation (\ref{upper_bound}) is valid for uniformly hyperbolic systems
of the class for which local quantities provide good approximations
for global quantities (Tent map, for example). For the other types of
uniformly hyperbolic systems and the non-uniformly hyperbolic systems
(as the Logistic map, H\'enon map, and Chua's circuit), the inequality
in Eq. (\ref{upper_bound}) should approximate well the divergence on
initial conditions for trajectories departing from the considered
region. Therefore, to obtain a good estimation of the Lyapunov
exponent one should consider an average of this quantity taken over
many regions in phase space as in

\begin{equation}
\langle H_{KS} (\tau) \rangle \leq \frac{1}{L}
\sum_{k=1}^{L}\frac{1}{\tau}\log{\left[\frac{N_{NR}(\tau,{\mathcal{B}})}{\rho(\tau,{\mathcal{B}}_k)}\right]}
\label{upper_bound2}
\end{equation}
\noindent
where we are considering that this average is taken over
${\mathcal{B}}_k$ regions in phase space, with $k=1,\ldots,L$.

Similarly, for experimental systems, we use 
\begin{equation}
\langle H_{KS} (\tau) \rangle \leq \frac{1}{L}
\sum_{k=1}^{L}\frac{1}{\tau}\log{\left[\frac{1}{\rho(\tau,{\mathcal{B}}_k)}\right]}
\label{upper_bound3}
\end{equation}

Let us now compare our result in Eq. (\ref{upper_bound}) with a
rigorous result \cite{saussol1} valid for a chaotic uniformly
hyperbolic map (piecewise monotone maps of an interval with a finite
number of branches and with bounded derivative of $p$-bounded
variation with an invariant measures with positive entropy) that
presents one positive Lyapunov exponent.  For $\epsilon \rightarrow
0$, we have that $\epsilon
\approxeq  \mu({\mathcal{B}})$. From Kac's lemma we have that 
$\frac{1}{\langle \tau \rangle}=\mu({\mathcal{B}})$ and then $\epsilon
\approxeq \frac{1}{\langle
\tau \rangle}$.  Also, for a sufficiently 
small interval, $\rho(\tau_{min},{\mathcal{B}}) \approxeq
\mu({\mathcal{B}})$ (assuming that $\rho(\tau,{\mathcal{B}})$ is well
described by an exponential distribution), then
Eq. (\ref{upper_bound1}) can be rewritten as $\lambda
\cong 
\frac{-\ln{(\epsilon)}}{\tau_{min}}$, which
agrees with the result derived in Ref. \cite{saussol1} that $\lambda =
\frac{-\ln{(\epsilon)}}{\tau_{min}}$, for $\tau_{min} \rightarrow \infty$. 

\section{Correlation function for short-term returns}\label{secao_correlacao}

The distribution $\rho(\tau,{\mathcal{B}})$ can be used to calculate
the correlation function, a quantity that measures how fast a system
looses information about the past as it evolves. Calling $\phi(x)$ to
be some observable measured at the position $x$, the correlation
between the observable $\phi$ at a time $T$ [i.e., $\phi(F^T(x))$]
with the initial observation $\phi(x)$ is given by \cite{stocastic}
\begin{eqnarray}
C(\phi,T) &=& \int \phi(x)\phi(F^T(x))d\mu \label{correlation} \\
- && \int \phi(x) d\mu.\int
\phi(F^T(x)) d\mu \nonumber
\end{eqnarray}
\noindent
where $d\mu$ stands as usually to $\sigma(x) dx$ and $\sigma(x)$
represents the invariant density from which the invariant measure
$\mu({\mathcal{B}})$ can be calculated by $\mu({\mathcal{B}}) =
\int_{x \in \mathcal{B}} \sigma(x)dx$

In this work, we are mainly interested in understanding the behavior
of the PRTs to regions ${\mathcal{B}}$. So, instead of averaging the
correlation of trajectories over the whole space $x$, we calculate the
correlation of trajectories in ${\mathcal{B}}$. Employing similar
ideas to the ones in Refs. \cite{correlation1,correlation2,paulo}, and
writing the observable to be $\mu({\mathcal{B}})$, we have that
\begin{equation}
C(\tau,{\mathcal{B}}) = \mu[{\mathcal{B}} \cap F^{-\tau}({\mathcal{B}})] -
\mu[{\mathcal{B}}]
\mu[F^{-\tau}({\mathcal{B}})]
\label{correlation11}
\end{equation}
\noindent
where $C(\tau,{\mathcal{B}})$ measures the correlation between
trajectories that visit the region ${\mathcal{B}}$ and that return to
it after $\tau$ iterations.  This function is also known as the speed
of mixing.

For an invariant measure, we have that $\mu[{\mathcal{B}}]=
\mu[F^{-\tau}({\mathcal{B}})]$, and thus, 
\begin{equation}
C(\tau,{\mathcal{B}})=\mu[{\mathcal{B}} \cap
F^{-\tau}({\mathcal{B}})]-\mu[{\mathcal{B}}]^2.
\label{correlation22}
\end{equation} 

The quantity $\mu[{\mathcal{B}} \cap F^{-\tau}({\mathcal{B}})]$
represents the probability measure of having trajectory points leaving
${\mathcal{B}}$ and returning to it after a series of returns
$\tau_i$, with $i=1,\ldots,l$ such that $\sum_i^l \tau_i = \tau$ and
$\tau_i \leq
\tau$. We can write the set $[{\mathcal{B}} \cap F^{-\tau}({\mathcal{B}})]$ 
as a union of two sets $S^{\prime} \cup S^{*}$ with $S^{\prime} \cap
S^{*} = \emptyset$, where $S^{\prime}$ as defined in Eq. (\ref{mu_NR})
and $S^*$ defined as the set of points that are mapped to
${\mathcal{B}}$ after $\tau$ iterations but with the additional fact
that for each $x^* \in S^*$ {\bf it exists} $\tau^* < \tau$ for which
$F^{\tau^*}(x^*) \in {\mathcal{B}}$. In other words, $S^{*}$
represents the set of points that are mapped back to ${\mathcal{B}}$
after $\tau$ iterations, excluding all the points that firstly return
to ${\mathcal{B}}$ after $\tau$ iterations.
 
If $\tau < 2\tau_{min}$, then, $\mu[{\mathcal{B}} \cap
F^{\tau}({\mathcal{B}})] = \mu(S^{\prime})$. Using Eqs. (\ref{mumu})
and (\ref{mu_NR}), we arrive to that $\mu[{\mathcal{B}} \cap
F^{\tau}({\mathcal{B}})]=
\mu({\mathcal{B}})
\rho(\tau,{\mathcal{B}})$, since there cannot be trajectory points that 
return more than once within this time interval.

Then, for $\tau < 2\tau_{min}$
\begin{equation}
C(\tau,{\mathcal{B}}) =
\mu({\mathcal{B}})\rho(\tau,{\mathcal{B}})-\mu[{\mathcal{B}}]^2
\label{correlation4}
\end{equation}

For $\tau  >  2\tau_{min}$, we can always write that  
\begin{equation}
C(\tau,{\mathcal{B}}) \leq
\mu({\mathcal{B}})\rho(\tau,{\mathcal{B}})-\mu[{\mathcal{B}}]^2, 
\label{correlation3}
\end{equation}
\noindent
since $\mu[{\mathcal{B}} \cap F^{\tau}({\mathcal{B}})]$ =
$\rho(\tau,{\mathcal{B}})
\mu({\mathcal{B}}) +\mu[S^*]$, 
Notice that as $\tau$ grows, $\mu[S^*] \rightarrow \mu_{R}$ in
Eq. (\ref{eq_celso1}).

Therefore, if the density of the returns has an exponential decay with
respect to time, so will the correlation behave, a statement
that was rigorously proved for some classes of systems and observables in
Ref. \cite{young}. The advantage of the approach proposed here is 
the simplicity with which we can understand such a rigorous result.
  
From Eqs. (\ref{correlation4}) and (\ref{correlation3}) we can clearly
see that the larger $(\rho(\tau,{\mathcal{B}}) - \mu[{\mathcal{B}}])$ is,
the slower the decay of the correlation function. That is exactly the
case for type II and III non-hyperbolic regions, for $\tau <
2\tau_{min}$, when the distribution $\rho(\tau,{\mathcal{B}})$
receives a large contribution of the power-law term $\beta_R$. For the
type I hyperbolic regions, $C(\tau,{\mathcal{B}})$ decays much faster
to 0, since $\rho(\tau,{\mathcal{B}}) \approx
\mu({\mathcal{B}})$.  And typically, the bigger $\omega$ in Eq. (\ref{dynamical_component}) is, 
the bigger $(\rho(\tau,{\mathcal{B}}) - \mu[{\mathcal{B}}])$. That is so
because the smaller $\tau_{min}$ is, the larger
$\rho(\tau_{min},{\mathcal{B}})$. 

In Ref. \cite{paulo}, a similar derivation of the correlation function
was proposed, but considering the correlation between trajectories
departing from $S^{\prime}$ and arriving to ${\mathcal{B}}$. In other
words, the correlation between trajectories that produce first returns
and that return only once to ${\mathcal{B}}$. In
Eqs. (\ref{correlation4}) and (\ref{correlation3}), one can estimate
the correlation between all trajectories departing from
${\mathcal{B}}$ and arriving in ${\mathcal{B}}$ including also
trajectories that return more than once.

\section{Determinism and stochasticism in the H\'enon map and the experimental Chua's circuit}\label{secao_aplicacao}

\begin{figure}[!htb]
\centerline{\hbox{\psfig{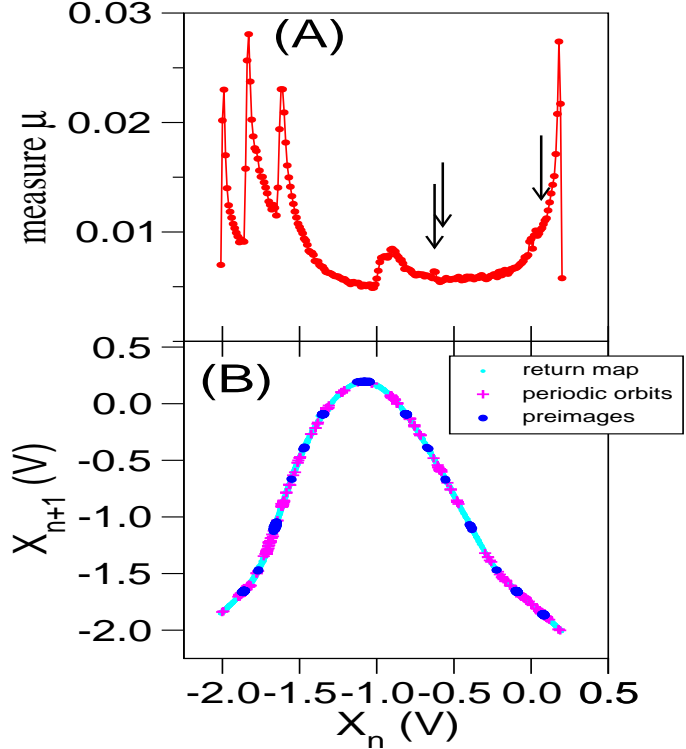}}}
\caption{[Color online] Results from the experimental Chua's circuit. 
(A) Probability measure with which $X_n$ visits intervals of length
$\epsilon=0.005V$. The arrows (from left to right) indicate the
intervals $\mathcal{B}_1$, $\mathcal{B}_3$, and $\mathcal{B}_2$,
respectively. (B) First returning map $X_n \times X_{n+1}$ in full
gray squares, pre-images of trajectory points located at the maximum
of the returning map in stars, and the UPOs of up to period 6 in
pluses.}
\label{medida_e_mapa01}
\end{figure}

\begin{figure}[!htb]
\centerline{\hbox{\psfig{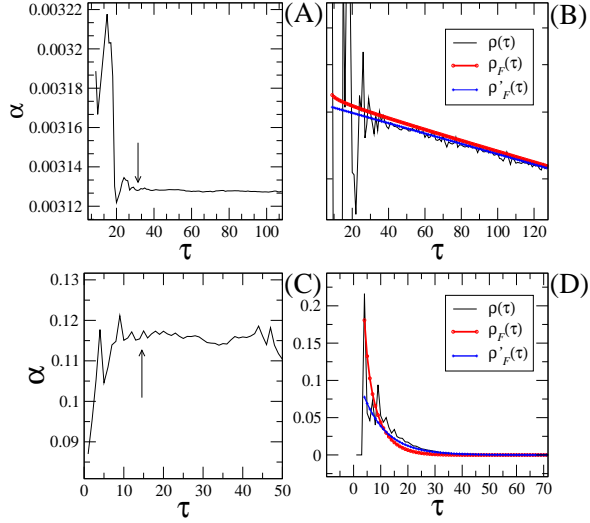}}}
\caption{[Color online] (A) and (B) are results for the H\'enon attractor and
(C) and (D) results for the experimental Chua's circuit. Estimation of
the parameter $\alpha$ and $\tau^{min}_{UPO}$ for the H\'enon
attractor (A) and for the Chua's circuit (C). In (B) and (D), we show
$\rho(\tau,{\mathcal{B}})$, the fitting of $\rho(\tau,{\mathcal{B}})$
by a function of the type $\rho_F(\tau,{\mathcal{B}})$
[Eq. (\ref{return_dist1})], and the fitting of
$\rho(\tau,{\mathcal{B}})$ by an exponential function of the type
$\rho_F^{\prime}(\tau,{\mathcal{B}})$ [Eq. (\ref{return_dist})]. The
fittings are made by considering the transformed time $\tau^{\prime}$
[Eq. (\ref{tau_prime})] but, in these figures, we re-transform the time
back to $\tau$ in order to plot the fittings together with the distribution
$\rho(\tau,{\mathcal{B}})$.}
\label{figura1}
\end{figure}

The H\'enon's map is given by $x_{n+1}=a-x_n^2+by_n$, and
$y_{n+1}=x_n$, with $a=1.4$ and b=$0.3$, and the considered
experimental Chua's circuit can be seen in
Ref. \cite{maranhao_PRE2008}. We focus our analyzes in three regions
$\mathcal{B}_1$ (type I), $\mathcal{B}_2$ (type II), and
$\mathcal{B}_3$ (type III).

For the H\'enon attractor, the regions ${\mathcal{B}}$ are boxes of
equal sides $2\epsilon$. The region $\mathcal{B}_1$ represents a box
centered at $(x,y)=(1.11807,0.14719)$, $\mathcal{B}_2$ a box centered
at the primary homoclinic tangent $(x,y)=(1.780098,0.09495)$, and
$\mathcal{B}_3$ a box centered at a period-2 UPO
$(x,y)=(1.36612008,-0.666120078)$.

\begin{figure}[!htb]
\centerline{\hbox{\psfig{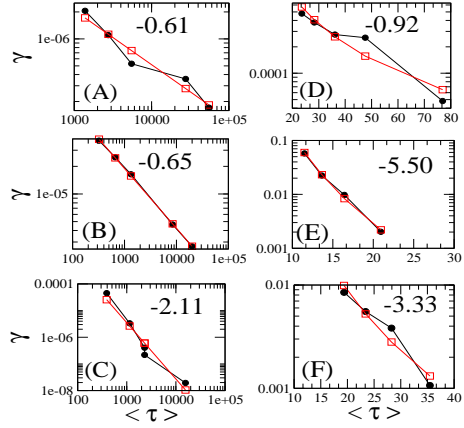}}}
\caption{The fittings of $\gamma$ (vertical axis) versus 
$\langle \tau \rangle$ in a log-log graph, for the distributions
$\rho_F$ of H\'enon map in (A-C) (left column) and for the
experimental Chua's circuit in (D-F) (right column), in the form
$\gamma \propto
\langle \tau \rangle ^{\theta}$ [see Eq. (\ref{teta})]. 
The value of the power $\theta$ is shown inside the
figures. Experimentally, the value of $\langle \tau \rangle$ is
over-estimated, which leads to a smaller $\theta$ exponent. 
}
\label{figura2}
\end{figure}

For the Chua's circuit, we reconstruct the attractor using the same
techniques of Ref. \cite{maranhao_PRE2008} and make the
trajectory discrete, producing a discrete time series represented by $X_n$,
which represents the voltage in the capacitor $C_1$, whenever the
voltage in the capacitor $C_2$ reaches zero.  The attractor is of the
R\"ossler-type. In Fig. \ref{medida_e_mapa01}(A) we show the invariant
measure of $X_n$ and in \ref{medida_e_mapa01}(B), the first returning
map $X_n \times X_{n+1}$, by the gray full squares.  The (blue) stars
represent trajectory points located at the maximum of the return map
and its pre-images. Every pre-image is located at a $X_n$ point for
which the probability density is large in (A), what typically happens
for non-uniformly hyperbolic systems. The plus symbol indicates the
place of the lower-period UPOs (up to period 6). In this circuit,
${\mathcal{B}}$ represents intervals of size
$\epsilon$. $\mathcal{B}_1$ represents the interval centered at
$X_n=-0.64$, $\mathcal{B}_2$ the interval centered at $X_n=0.05$, and
$\mathcal{B}_3$ the interval centered at the position of a period-2
UPO $X_n=-0.586637$. These points where the intervals are positioned
are indicated by the arrows in Fig. \ref{medida_e_mapa01}(A).

We now make a detailed analysis of the PRTs to the box
$\mathcal{B}_2$, for the H\'enon map [in Figs. \ref{figura1}(A-B)],
and the PRTs to the interval $\mathcal{B}_2$, for the Chua's circuit
[in Figs. \ref{figura1}(C-D)]. In (A) and (C), we estimate the value
of $\alpha$ as we consider different values for $\tau^{min}_{UPO}$, in
Eq. (\ref{estima_alpha}). We consider the smallest value of
$\tau^{min}_{UPO}$ for which the value of $\alpha$ "converges".
The arrow indicates these values. For the H\'enon attractor, $\tau^{min}_{UPO}=28$
and for the Chua's circuit, $\tau^{min}_{UPO}=13$.  In (B) and (D), we show
$\rho(\tau,{\mathcal{B}})$, $\rho_F(\tau,{\mathcal{B}})$, and a fitting of $\rho(\tau,{\mathcal{B}})$ by an exponential
function of the type of ${\rho}^{\prime}_F(\tau,{\mathcal{B}})$
[Eq. (\ref{return_dist})]. For the H\'enon attractor (B), using the
distribution $\rho_F(\tau,{\mathcal{B}})$ in the integral of Eq. (\ref{cond1}) produces 
the value 0.994826 while using $\rho_F^{\prime}(\tau,{\mathcal{B}})$ produces 
0.987701. For the Chua's circuit (D), using the distribution $\rho_F(\tau,{\mathcal{B}})$ in
the integral of Eq. (\ref{cond1}) produces the value 0.78 while using
$\rho_F^{\prime}(\tau,{\mathcal{B}})$ produces 0.68. Reminding that the integrals in
Eq. (\ref{cond1}) should provide 1, it is clear that the proposed distribution in
Eq. (\ref{return_dist1}) fits better $\rho(\tau,{\mathcal{B}})$.

In Fig. \ref{figura2}, we show the relation between $\gamma$ and
$\langle \tau \rangle$ for $\mathcal{B}_1$, $\mathcal{B}_2$, and
$\mathcal{B}_3$, in the H\'enon attractor [first column, (A-C)] and in
the Chua's circuit [second column, (D-F)]. We have numerically
obtained that for the H\'enon map, $\gamma \propto
\langle \tau \rangle^{\theta}$ with $\theta \cong -0.6$ 
[excluding the region ${\mathcal{B}}_3$, whose results are shown in (C)], and
for the Chua's circuit $\gamma
\propto \langle \tau \rangle^{\theta}$ with $\theta < -0.9$. 
Remind that the larger is $|\theta|$, the smaller $\beta_R$.

For type II non-hyperbolic regions, as $\epsilon$ is decreased,
typically we expect that the unstable manifold of the UPO outside the
region will still belong to the region, which leads to a small
$|\theta|$ value.

\begin{table*}
\caption{Estimation of the Lyapunov exponent. For the H\'enon map, we typically find
that the UPO with the smallest period has a period larger than the
observed $\tau_{min}$, a consequence of the non-hyperbolic character
of this map [see Fig. \ref{dariel_figura3}(B) and
\ref{dariel_figura3}(D)]. Thus, for calculating the Lyapunov exponent,
we consider in Eq. (\ref{upper_bound}) $\tau=P_{min}$, where $P_{min}$
is the lowest period of all UPOs inside ${\mathcal{B}}$. For all
regions studied in this map, the number of non-recurrent UPOs is
within the interval $N_{NR}=[1,2]$ as expected. For the experimental
Chua's circuit, we calculate the Lyapunov exponent considering in
Eq. (\ref{upper_bound1}) $\tau$ equal to the first Poincar\'e return
for which the PDF presents its third maximum. We also assume that
$N_{NR}=1$. The positive Lyapunov exponent of the H\'enon attractor is
also calculated using the technique in Ref. \cite{eckmann} and the
largest Lyapunov exponent of the Chua's circuit is also calculated
using the technique in Ref. \cite{sano}, with the code of the Tisean
package
\cite{tisean}.  We obtain that the positive
Lyapunov exponent of the H\'enon map is 0.419/iteration and the one
for the Chua's circuit is 0.52/cycle. The value of $\tau$ used
is the number between parentheses.}
\label{tabela1}
\begin{tabular}{c|c|c|c|c|c}
\hline
$\epsilon$ - H\'enon & 0.02 & 0.01 & 0.005 & 0.001 & 0.0008 \\
\hline
${\mathcal{B}}_1$ & 0.32(21) & 0.40(23) & 0.40(23) & 0.39(23) & 0.39(23) \\
${\mathcal{B}}_2$ & 0.61(18) & 0.33(18) & 0.36(18) & 0.40(22) & 0.29(29) \\
\hline
\hline
$\epsilon$ - Chua & 0.1700 & 0.1450 & 0.1200 & 0.0950 & 0.0700 \\
\hline
${\mathcal{B}}_1$ & 0.43(7) & 0.42(7) & 0.42(7) & 0.42(7) & 0.46(9) \\
${\mathcal{B}}_2$ & 0.58(5) & 0.39(7) & 0.58(5) & 0.25(13) & 0.27(13)
\\
\hline
\end{tabular}
\end{table*}

Table \ref{tabela1} shows estimates for the positive Lyapunov exponent
using the right-hand side of Eq. (\ref{upper_bound}) for the H\'enon
map and Eq. (\ref{upper_bound1}) for the Chua's
circuit. ${\mathcal{B}_1}$ represents a hyperbolic type I region and
${\mathcal{B}_2}$ a non-hyperbolic type III region, with different
sizes $\epsilon$.  For the H\'enon attractor, a numerical calculation
of this exponent provides $\lambda$=0.419
\cite{eckmann}, and for the experimental Chua's circuit, $\lambda$=0.52/cycle
\cite{sano,tisean}. The unit of the Lyapunov exponent for this circuit 
is in [1/cycles]. That is due to the fact that for the calculation of
this exponent we have used the discrete time series $X_n$, which
represents the voltage in capacitor $C_1$ whenever the voltage of
capacitor $C_2$ is zero.

In both systems, the Lyapunov exponents estimated from
Eqs. (\ref{upper_bound}) [for the H\'enon map] and
(\ref{upper_bound1}) [for the Chua's circuit] using returns measured
in the non-hyperbolic regions ${\mathcal{B}}_2$ produce worse
estimates than the ones measured in the hyperbolic regions
${\mathcal{B}}_1$. That is to be expected since our estimations are
valid for uniformly hyperbolic systems. For both systems and all
regions, the estimation of the Lyapunov exponents produce better
results as the size of the regions decrease and $\tau_{min}$ becomes
large. That is also to be expected since as the size of the box
decreases the chance that a region has a hyperbolic character
increases.

Since both the H\'enon map and the Chua's circuit are non-uniformly
hyperbolic systems, in order to obtain better estimates for the
Lyapunov exponents we need to calculate an average value considering
many regions in phase space.

To firstly illustrate how an average value provides a better
estimation for the real value of the Lyapunov exponent, we average all
the values shown in Table \ref{tabela1}. For the H\'enon map, we
obtain that $\lambda=0.39$ (the real Lyapunov exponent is 0.419) and
for the Chua's circuit we obtain that $\lambda=0.42$ (the real value
is 0.52).

Then, we calculate $\langle H_{KS} \rangle$ for the H\'enon map, using
Eq. (\ref{upper_bound2}) and considering $L=$500 regions with
$\epsilon$=0.005 and $\tau=P_{min}$, where $P_{min}$ is the lowest
period of all UPOs inside each region. We obtain that $\langle H_{KS}
\rangle=0.464$ (compare with the real value 0.419). These $L=$500
regions are centered in points of a 500 long trajectory. Using
Eq. (\ref{upper_bound3}) for the same previous conditions, we obtain
$\langle H_{KS}
\rangle=0.441$. For the Chua's circuit, we calculate $\langle H_{KS}
\rangle$ using Eq. (\ref{upper_bound3}) considering $L=$25 regions
${\mathcal{B}}$ with $\epsilon$=0.067 and $\tau$ equal to the second
largest first Poincar\'e return. We obtain that $\langle H_{KS}
\rangle=0.42 \pm 0.1$ (compare with the real value 0.52).

\begin{figure}[!htb]
\centerline{\hbox{\psfig{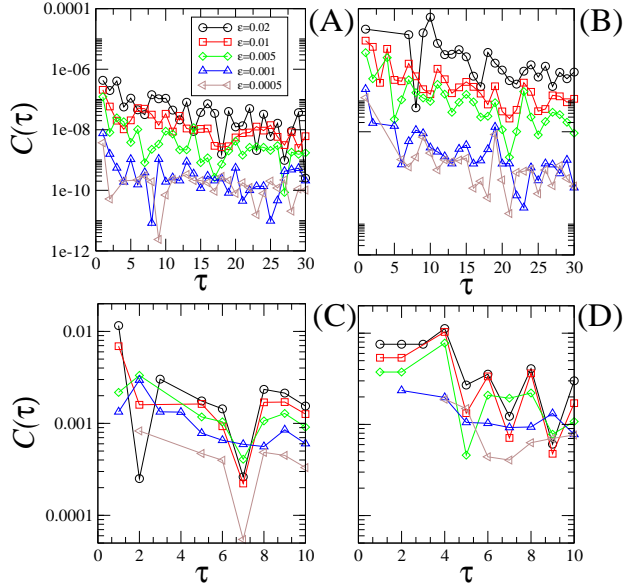}}}
\caption{Correlation function with respect to $\tau$ for the H\'enon map (A-B)
and the experimental Chua's circuit (C-D). In (A) and (C), we consider
the type I hyperbolic regions ${\mathcal{B}}_1$, and in (B) and (D),
we consider the type II non-hyperbolic regions ${\mathcal{B}}_2$.}
\label{fig_correlacao}
\end{figure}

In Fig. \ref{fig_correlacao}, we show the decay of the correlation
function with respect to $\tau$. Notice that this function is local
and reflects the decay of correlation of PRTs to a particular
interval. As expected, the correlation function decays faster in the
type I hyperbolic regions [(A) and (C)] than in the type II
non-hyperbolic regions. In addition, in general, the larger $\epsilon$
is, the larger the value of the correlation function, a consequence of
the fact that the larger $\epsilon$ is, the larger both $\theta$ and
$\omega$ are.

\section{Conclusions}\label{secao_conclusao}

We have studied the Poincar\'e recurrence time (PRT) of chaotic
systems, a quantity that measures the time interval between two
consecutive visits of a trajectory to a finite size region in phase
space. The motivation in studying PRT is that for some systems the
only possible measure one can make is the time interval between two
events. But one still wants to understand what kind of dynamics is
behind the generation of these returns.

If the region in phase space has an arbitrarily small size, for
systems that have strong mixing characteristics, a long return time
contains no longer information about the future returns, which means
that the consecutive series of returns lose their correlation and they
behave as if they had been generated by a completely random
process. At such a situation, the probability distribution of the PRTs
approaches a Poisson distribution
\cite{collet,hirata} and few can be said about the dynamical manifestations of the
data by only considering the form of this distribution.

Our approach is mainly devoted to characterize a chaotic system
(obtaining relevant invariant quantities) considering regions with
finite size and short return times. We show how to calculate the
short-term correlation function [see Eq. (\ref{correlation4})] and the
Kolmogorov-Sinai entropy [see Eqs. (\ref{upper_bound1}) and
(\ref{upper_bound3})], using the distribution of PRTs, a quantity that
can be easily accessible in experiments.

As the region in phase space where the returns are being measured
becomes larger, the first Poincar\'e returns reflect more the
deterministic nature of the system. That leads to a large deviation of
the density of the returns from the Poisson [see
Eq. (\ref{return_dist1})] and a slower decay of correlations for
short-term returns [see Eq. (\ref{correlation4}) and
Fig. \ref{fig_correlacao}].

These characteristics can be advantageously used to characterize the
level of determinism in chaotic systems. More specifically, the larger
$\theta$ is in Eq. (\ref{dynamical_component}), the more deterministic
is the considered chaotic system under the point of view of the
PRTs. Another quantity is $\omega=\tau^{min}_{UPO} -
\tau_{min}$, where $\tau^{min}_{UPO}$ is the return value for which the
probability distribution of the PRTs presents a continuous decay with
the increasing of the return time, and $\tau_{min}$ is the return with
the minimal length. The larger $\omega$ is, the larger is the time
span within which discontinuities are observed in the distribution of
returns. Each discontinuity identifies particular returns, which makes
them to be highly predictable.  As shown in
Sec. \ref{secao_correlacao}, these previous intuitive ideas are indeed
correct, i.e., larger $\theta$ and $\omega$ result in a slower decay
of correlations.

The strategy of using a value of $\tau$ that is larger than
$\tau_{min}$ in Eqs. (\ref{upper_bound2}) and (\ref{upper_bound3}) is
due to the fact that in non-hyperbolic systems as the H\'enon map and
the experimental Chua's circuit the density of the PRTs for
$\tau=\tau_{min}$, i.e. $\rho(\tau_{min},{\mathcal{B}})$, receives a
contribution coming from UPOs outside of ${\mathcal{B}}$, which
violates the conditions under which these equations are derived. When
the UPOs are known as it is the case for the H\'enon map, we use
$\tau=P_{min}$, in Eqs. (\ref{upper_bound2}) and (\ref{upper_bound3}),
where $P_{min}$ represents the period of the UPO inside
${\mathcal{B}}$ with the lowest period. When the UPOs are unknown as
it is the case for the experimental Chua's circuit, we use $\tau >
\tau_{min}$.

Our approach can be straightforward applied to the treatment of data
coming from complex systems, and the reason lies on the formulas that
we have used. As one can check the 3 most relevant formulas used in
our work [Eqs. (\ref{dynamical_component}),
(\ref{upper_bound3}), and (\ref{correlation4})] depend
only on the probability of the Poincar\'e return times to finite size
intervals with short time-length, a quantity that can be easily and
quickly accessible from measurements and that, in principle, does not
require the existence of chaos.

For some technological applications as in the control of chaos
\cite{OGY}, 
one does not need to know with high precision the position of an UPO,
but rather its unstable eigenvalue. Equation (\ref{rho_Lmin}) offers a
trivial way to calculate this quantity by measuring the probability
with which the fastest first Poincar\'e return to a sufficiently small
interval happens. Naturally, this equations also offers a simple way
to obtain estimations for the local first derivative of a system, a
quantity often needed to make local models of complex systems.  If the
system is higher-dimensional then $L_{min}$ in Eq. (\ref{rho_Lmin})
should refer to the product of all the unstable eigenvalues of a
single non-recurrent UPO appearing in the region, as explained in
Ref. \cite{celso}.

{\bf Acknowledgments: } MSB would like to thank the wonderful
discussions with Benoit Saussol, Miguel Mendes, and Tomas Persson,
during the International conference in honor of Yakov Pesin on his
60th birthday (25-29 June, 2007), concerning the Poincar\'e return
time to regions of finite size, discussions which partially influenced
the many ideas presented in this work, and also discussions with
M. Todd about recent mathematical proofs concerning the exponential
statistics of the PRTs.  MSB was also partially supported by the
Centro de Matem\'atica da Universidade do Porto, financed by FCT
through the programmes POCTI and POSI, with Portuguese and European
Community structural funds.  JCS and DMM acknowledge the financial
support of CNPq and FAPESP.


\appendix

\section{The density of returns $\rho$ and the non-recurrent UPOs}\label{apendice}

In Ref. \cite{celso} it is derived a formula that relates the
probability measure of a D-dimensional box ${\mathcal{B}}$ with the
unstable eigenvalues of the UPOs inside it. More specifically,
\begin{equation}
\mu({\mathcal{B}})=\lim_{P \rightarrow \infty}\left[ \sum_j L_j(P)^{-1} \right]
\label{eq_celso}
\end{equation}
\noindent
where $L_j(P)=L_1j(P) L_2j(P) \ldots L_Uj(P)$, and $L_uj(P)$
($u=1,\ldots,U$) represent the $U$ positive eigenvalues (larger than
1) of the $j$ fixed point located in ${\mathcal{B}}$ of the $P$-fold
iterate of the map represented by  $F^P$ (i.e., the fixed points are period-P UPOs that
belongs to ${\mathcal{B}}$), where $P$ tends to infinity. In a general
situation, there are $U$ positive eigenvalues and the box is
$D$-dimensional. In the following, we consider $D$=2, and U=1 (there
is only one positive Lyapunov exponent).

Equation (\ref{eq_celso}) was demonstrated to hold for mixing hyperbolic 
(axiom A) attractors and was shown numerically to hold for the non-hyperbolic
H\'enon attractor in Refs.
\cite{lai,baptista:2005}, for UPOs of moderately large period $P \cong 30$.

As done in Ref. \cite{paulo}, we rewrite the right-hand side of
Eq. (\ref{eq_celso}) as a sum of two terms, without taking the limit
of $P \rightarrow \infty$ but for a finite $P$:
\begin{equation}
\sum_j L_j(P)^{-1}  = \mu_{R} + \mu_{NR}
\label{eq_celso1}
\end{equation}
\noindent
\noindent
where $\mu_{R} = \sum_j [L^{R}_j(P)]^{-1}$ and 
\begin{equation}
\mu_{NR} = \sum_j [L^{NR}_j(P)]^{-1}, 
\label{mu_NR1}
\end{equation}
\noindent
where $L^{R}_j(P)$ are the unstable eigenvalues of the so called {\it
recurrent} UPOs that return to ${\mathcal{B}}$ more than once before
completing their cycles, and $L^{NR}_j(P)$ are the unstable
eigenvalues of the so called {\it non-recurrent} UPOs that return to
${\mathcal{B}}$ only once. So, while $\mu_{NR}$ measures the
contribution to the measure due to chaotic trajectories associated
with non-recurrent UPOs, $\mu_{R}$ measures the contribution to the
measure due to chaotic trajectories associated with recurrent UPOs.

As shown in Ref. \cite{paulo}, there is a clever way to relate the
density $\rho(\tau,{\mathcal{B}})$ with the measure of the attractor
associated with UPOs inside ${\mathcal{B}}$ by
\begin{equation}
\rho(\tau,{\mathcal{B}})=\mu_{NR}(\tau,{\mathcal{B}})
\label{mumu}
\end{equation}

The term $\mu_{NR}(\tau,{\mathcal{B}})$ can be represented in terms of
space averages by
\begin{equation}
\mu_{NR}(\tau,{\mathcal{B}})=\frac{\mu(S^{\prime})}{\mu({\mathcal{B}})}
\label{mu_NR}
\end{equation}
\noindent
where $\mu(S^{\prime})$ represents the measure of the set $S^{\prime}$
(the part of the measure inside ${\mathcal{B}}$ due to the set
$S^{\prime}$) and $S^{\prime}$ represents the set of points of the
attractor that returns firstly to ${\mathcal{B}}$ after $\tau$
iterations. More rigorously, representing by $F$ the transformation
that generates a chaotic set $A$ and given a subset ${\mathcal{B}}
\subset A$, then $S^{\prime} \in A$ and $S^{\prime} =
F^{-\tau}({\mathcal{B}}) \cap {\mathcal{B}}$ such that there is not
$\tau^{*}<\tau$ for which $F^{\tau^{*}}({\mathcal{B}}) \cap {\mathcal{B}}
\neq \emptyset$. 

But the right-hand side of Eq. (\ref{mu_NR}) can be estimated by
$\frac{N_{NR}(\tau,{\mathcal{B}})}{N(\tau)}$, 
and therefore, 
\begin{equation}
\mu_{NR}(\tau,{\mathcal{B}})=\frac{N_{NR}(\tau,{\mathcal{B}})}{N(\tau)}, 
\end{equation}
\noindent
where
$N_{NR}(\tau,{\mathcal{B}})$ and $N(\tau)$ represent the number of
non-recurrent UPOs with period $\tau$ inside ${\mathcal{B}}$ and
$N(\tau)$ the total number of different UPOs of period $\tau$ embedded
in the chaotic attractor.

Now, using Eq. (\ref{mu_NR1}), it is clear that 
 \begin{equation}
\rho(\tau,{\mathcal{B}}) \geq
L_{min}(\tau)^{-1},
\label{mumu1}
\end{equation}
\noindent
where $L_{min}$ represents the unstable eigenvalue with the lowest
amplitude within all UPOs with period $\tau$.

\begin{figure}[!htb]
\centerline{\hbox{\psfig{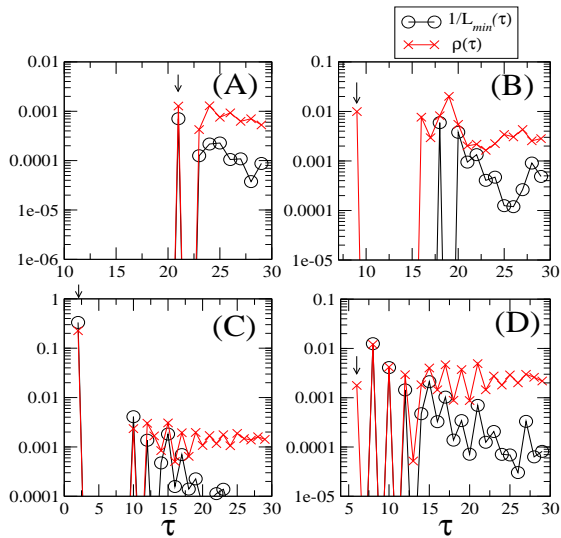}}}
\caption{By empty circles we show $1/L_{min}(\tau)$ and (red) crosses the value of 
$\rho(\tau,{\mathcal{B}})$, for the H\'enon attractor for different
regions. $\mathcal{B}_1$ [in (A)], $\mathcal{B}_2$ [in (B)], and
$\mathcal{B}_3$ [in (C)], and also for a region centered at a period-8
UPO [in (D)]. $\epsilon=0.02$ in all figures.  The arrow points to the
value of $\tau_{min}$.}
\label{dariel_figura3}
\end{figure}

In Fig. \ref{dariel_figura3}, we show the values of $
L_{min}(\tau)^{-1}$ and the density of returns $\rho$ in the H\'enon
attractor, for the regions $\mathcal{B}_1$ [in (A)], $\mathcal{B}_2$
[in (B)], $\mathcal{B}_3$ [in (C)], and also for a region centered at
a period-8 UPO [in (D)] at the position $(x,y)$=(1.496703, -0.545333),
denoted by ${\mathcal{B}}_4$.

We see that for $\tau \approxeq \tau_{min}$, inequality in
Eq. (\ref{mumu1}) is close to an equality and we can write that
\begin{equation}
\rho(\tau_{min},{\mathcal{B}}) \approxeq L_{min}(\tau_{min})^{-1}.
\label{rho_Lmin}
\end{equation}

The reason is that for such a case, there is only a few (a number of
the order of 1) non-recurrent UPOs with period $\tau_{min}$. It is
easy to understand that by using the following argument. Imagine a
sufficiently small region centered around an UPO with period
$P=1$. Clearly $\tau_{min}=1$ and there will be only one non-recurrent
UPO with period $P$=1. Now, consider a region around an UPO with
period $P$=2. Similarly, $\tau_{min}=2$ and there will be only one UPO
with period $2$. Typically, for sufficiently small regions, there will
be only one non-recurrent UPO inside the regions if $P \approxeq
\tau_{min}$. 

Provided that the UPO is hyperbolic, the uniqueness of the UPO in the
small region is garanteed by the Hartman-Grobman Theorem
\cite{hartman} and the size of the region in which the uniqueness of
the UPO can be garanteed is related to the strength of the
hyperbolicity of the UPO.

\begin{figure}[!htb]
\centerline{\hbox{\psfig{file=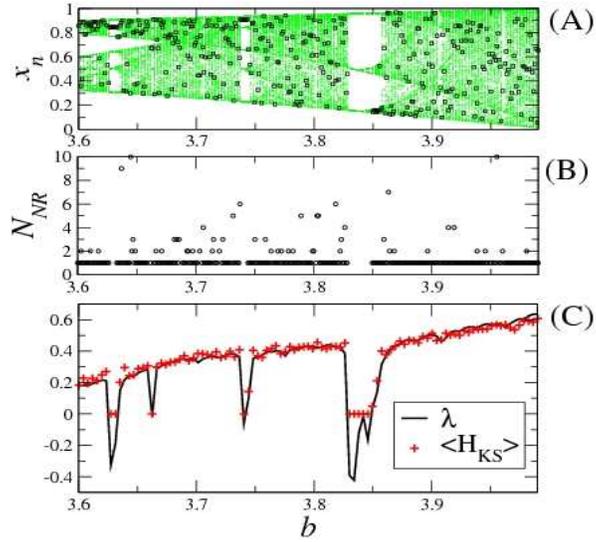,height=8cm,width=8cm,viewport=0 -20 450 470}}}
\caption{[Color online] Results for the Logistic map: $x_{n+1}=bx_n(1-x_n)$. 
(A) Bifurcation diagram of the Logistic map (green points). Empty
squares represent intervals of size $\epsilon=0.001$ randomly selected
such that $\tau_{min} \in [10,14]$. We consider 100 parameter values
within the range $[3.6,3.99]$ and for each parameter we consider one
interval. (B) the number $N_{NR}$ of non-recurrent UPOs with period
$P=\tau_{min}$ inside each one of the 100 intervals. (C) The value of
the Lyapunov exponent (black tick line), denoted by $\lambda$, and
$\langle H_{KS} \rangle$ (red pluses) calculated from
Eq. (\ref{upper_bound3}).}
\label{dariel_figura6}
\end{figure}

To illustrate that, we consider the Logistic map
$[x_{n+1}=bx_n(1-x_n)]$, whose bifurcation diagram is shown in
Fig. \ref{dariel_figura6}(A), constructed considering 100 parameter
values $b$ within the parameter range $[3.6,3.99]$. In
Fig. \ref{dariel_figura6}(B), we show the number of non-recurrent
UPOs, denoted by $N_{NR}$, for UPOs with period $P=\tau_{min}$ and
intervals with size $\epsilon=0.001$, randomly selected such that
$\tau_{min} \in [10,14]$. For the large majority of intervals,
$N_{NR}=1$. Finally, in \ref{dariel_figura6}(C), we show the values of
$\langle H_{KS} \rangle$ calculated from Eq. (\ref{upper_bound3}) for
this parameter range. Notice that despite the choice of $N_{NR}=1$,
the value of $\langle H_{KS} \rangle$ is a good estimation for the
positive Lyapunov exponent of this map, indicated in this figure by
$\lambda$.

\begin{thebibliography}{99}

\bibitem{stocastic} M. Viana, Stochastic dynamics of deterministic systems, 
Brazillian Math. Colloquium IMPA, 1997; A. Lacosta, M. C. Mackey, Chaos,
Fractals, and noise, Springer, 1994.

\bibitem{dinamica} K. T. Alligood, T. D. Sauer, J. A. Yorke, Chaos, an
  introduction to dynamical systems, Springer, 1997; H. Kantz and T.
  Schreiber, Nonlinear time series analysis, Cambridge University
  Press, Cambridge, 1997.

\bibitem{mamen} C. Komalapriya, M. Thiel, M. C. Romano,  {\it et al.}, 
Reconstruction of a system's dynamics from short trajectories, Phys. Rev. E, 78 (2008) 
066217 (2008).

\bibitem{nicolis} G. Nicolis and C. Rouvas-Nicolis, in Encyclopedia of
Nonlinear Sciences, Routledge, New York, 2005.

\bibitem{poincare} H. Poincar\'e, Sur le probléme des trois corps et les \'equations de la dynamique, Acta Matematica, 13 (1890), 1-270.

\bibitem{paulo} P. R. F. Pinto, M. S. Baptista, I. Labouriau, First
  Poincar\'e return, natural measure, UPO and Kolmogorov-sinai
  Entropy, arXiv:0908.4575.


\bibitem{baptista} M. S. Baptista, I. L. Caldas, M. V. A. P. Heller, and A. A. Ferreira
Recurrence in plasma edge turbulence, Phys. Plasmas, 8 (2001), 4455-4462;
M. S. Baptista and I. L. Caldas, Stock Market Dynamics, Physica A, 312 (2002) 539-564.

\bibitem{grassberger} P. Grassberger and I. Procaccia, 
Estimation of the Kolmogorov entropy from a chaotic signal
Phys. Rev. A 28, (1983), 2591-2593.

\bibitem{cohen} A. Cohen and I. Procaccia, 
Computing the Kolmogorov entropy from time signals of dissipative and conservative dynamical systems, Phys. Rev. A, 31 (1985) 1872-1882.

\bibitem{correlation2} R. Artuso, Correlation decay and return time statistics, 
Physica D, 131 (1999), 68-77.

\bibitem{young} L.-S. Young, Recurrence times and rates of mixing,  
Israel Journal of Mathematics, 110, (1999), 153-188.

\bibitem{eduardo} Eduardo G. Altmann, Elton C. da Silva, and Iber\^e
  L. Caldas, Recurrence time statistics for finite size intervals,
  Chaos, 14 (2004), 975-981.

\bibitem{baptista:2005} M. S. Baptista, S. Kraut, C. Grebogi, 
Poincar\'e recurrence and measure of hyperbolic and nonhyperbolic
chaotic attractors, Phys. Rev. Lett., 95, (2005) 094101.

\bibitem{return} The Poincar\'e return time (PRT) is the number of iterations needed to make a
  trajectory leaving from a point in the attractor inside a region
  to return to it. A typical trajectory generates a series of returns
  $\tau_1$, $\tau_2, \ldots$, $\tau_L$.


\bibitem{hirata} B. Saussol, On fluctuations and the exponential
  statistics of return times Nonlinearity, 14 (2001), 179-191; M.
  Hirata, B. Saussol, and S. Vaienti, Statistics of return times: a
  general framework and new applications, Comm. Math. Phys. 206,
  (1999), 33-55.

\bibitem{eckmann} J.-P. Eckmann and D. Ruelle, Ergodic theory of chaos and strange attractors, Rev. Mod. Phys. 57 (1985) 617-656.  

\bibitem{saussol1} B. Saussol, Recurrence rate in rapidly mixing dynamical systems, 
Discrete and Continuous Dynamical Systems A, 15 (2006), 259-267.

\bibitem{correlation1} B. V. Chirikov, D. L. Shepelyansky, Correlation properties of dynamical chaos in Hamiltonian systems, Ninth International Conference of Nonlinear Oscillations, Kiev
1981, Naukova Dumka, Kiev, 1984, vol. II;  
B. V. Chirikov, D. L. Shepelyansky, 
Correlation properties of dynamical chaos in Hamiltonian systems,  
Physics D, 13, (1984) 395-400l; C. F. F. Karney,  Long-time correlations 
in the stochastic regime, Physica D, 8 (1983) 360-380.

\bibitem{maranhao_PRE2008} D. M. Maranh\~ao, M. S. Baptista, J. C. Sartorelli,
I. L. Caldas, Experimental observation of a complex periodic window,
Phys. Rev. E., 77 (2008) 037202.

\bibitem{sano} M. Sano and Y. Sawada, 
Measurement of the Lyapunov spectrum from a chaotic time series
Phys. Rev. Lett., 55 (1985) 1082-1085.  

\bibitem{tisean} R. Hegger, H. Kantz,
and T. Schreiber, Practical implementation of nonlinear time series
methods: the TISEAN package, Chaos, 9 (1999) 413.


\bibitem{collet} P. Collet, Some ergodic properties of maps of the interval,
Lectures given at the CIMPA summer school, Dynamical Systems and
Frustrated Systems, Hermann, Paris, 1996.

\bibitem{OGY} E. Ott, C. Grebogi, and J. A. Yorke, Controlling chaos, 
Phys. Rev. Lett. {\bf 64}, (1990) 1196-1199.

\bibitem{celso} C. Grebogi, E. Ott, J. A. Yorke, 
Unstable periodic orbits and the dimensions of multifractal chaotic
attractors, Phys. Rev. A, 37, (1988) 1711-1724.

\bibitem{lai} Y.-C. Lai, Y. Nagai, and C. Grebogi, 
Characterization of the Natural Measure by Unstable Periodic Orbits in
Chaotic Attractors, Phys. Rev. Lett. {\bf 79}, 649 (1997).

\bibitem{hartman} L. Perko, Differential Equations and Dynamical
Systems (Springer-Verlag, New York, 1991).

\end{thebibliography}
\end{document}